\documentclass[a4paper,12pt]{article}
\begin{document}

\title{Accelerated-like expansion: inhomogeneities versus dark energy}
\author{Marie-No\"elle C\'el\'erier\footnote{Electronic mail: marie-noelle.celerier@obspm.fr} \\
{\it \small LUTH, Observatoire de Paris-Meudon, 5 place Jules Janssen, 92195 Meudon Cedex, France}}
\date{\today}

\maketitle
\begin{abstract} 
The currently available cosmological data yield, as a most striking result, that the expansion rate of the universe seems to be increasing at late times, contrary to the standard (zero cosmological constant) FLRW prediction. The usual explanation for this discrepancy is that a new component of the energy density of the universe, known as dark energy, dominates this recent evolution. Since the existence of such a new component would have a revolutionary impact on our understanding of the fundamental laws of physics, we think important to check other interpretations. We have therefore shown that the SNIa observations could be reproduced by the effect of inhomogeneities. This idea has been further developed by different teams, and enlarged to other cosmological data. We will give here a review of the results of these works and the prospects for future developments. 
\end{abstract}

\section{Introduction}

The luminosity distance-redshift relation obtained from the observations of the SNIa, and interpreted with the a priori assumption that our Universe can be represented by an homogeneous Friedmann-Lema\^itre-Robertson-Walker (FLRW) model, implies that the expansion of the Universe is accelerating. This acceleration is generally ascribed to the influence of a dark energy component, a cosmological constant or a negative pressure fluid with $\Omega_\Lambda\sim 0.73$. However, a cosmological constant is usually interpreted as the vacuum energy of which current particle physics cannot explain the low amplitude and a negative pressure fluid remains a mysterious phenomenum. This is known as the cosmological constant problem. 

Another feature of this luminosity distance-redshift relation is to yield a late acceleration of the expansion rate, about the epoch when structure formation enters the non-linear r\'egime. This would imply that we live at a time when the matter density energy and the dark energy are of the same order of magnitude, which is known as the coincidence problem. 

To solve these problems, many works have been devoted to the understanding of the nature of a possible dark energy, others have explored the assumption that General Relativity could cease to be valid at large scales. 

Another more simple and natural proposal, which makes only use of known physics and phenomena, has been put forward seven years ago (C\'el\'erier 2000).  We have shown that the SNIa observations could be reproduced by the effect of inhomogeneities. This idea has been further developed by different teams, and enlarged to other cosmological data. We give here a review of the results of these works and prospects for future developments.

\section{Influence of inhomogeneities} 

\subsection{The first proposals}

Just after the release of the SNIa data, C\'el\'erier (2000) and 
independently Tomita (2000; 2001; 2003) have proposed that the apparent acceleration of the Universe expansion could be due to inhomogeneity effects. 

C\'el\'erier (2000) proceeds to an expansion of the $\Lambda = 0$ inhomogeneous luminosity distance-redshift relation up to the fourth order in powers of $z$ and shows that it can mimic the Friedmannian relation with a cosmological constant for $z<1$. She then proposes, as a mere example, a flat Lema\^itre-Tolman-Bondi (LTB) toy model with $\Lambda = 0$. Peculiar classes of this LTB toy model have been applied by Iguchi et al (2002) to the fitting of the $\Lambda$CDM model. Their attempt has been successful for $z<1$, but unsuccessful at large $z$, therefore giving a hint of a solution to the coincidence problem.

Tomita (2000; 2001; 2003) considers cosmological void models. A low-density inner homogeneous region is connected at $z_1\sim 0.067$ to another homogeneous region of higher-density. The observer located inside the void can be on-centre or off-centre. Both regions decelerate, but since the void expands faster than the outer region, an acceleration is experienced by the observer for $z_1<z<1.5$. 

Palle (2002) uses the averaging procedure of Buchert and Ehlers (1997) to study the effect of inhomogeneities on different cosmic observables. In the averaging language, the above void situation results in the simultaneous presence of largely under-dense and over-dense regions which can produce a large kinematical backreaction. 

\subsection{The revival}

Unfortunately, no much attention was at first payed to these proposals. But during the last two years, we have experienced a reniewed interest for these ideas. 

First, R\"as\"anen (2004) have studied backreaction phenomena with an averaged flat LTB as a toy model. He shows that, even for an exact solution, spatially averaging is not a covariant procedure and that the choice of the hypersurface on which this averaging is performed leads to different results. 

He then concludes that the two methods best appropriate for dealing with this issue are: either to proceed to a consistent second order perturbative analysis of a FLRW model; or to adopt a quasi-perturbative approach, i. e., a first order perturbative method for large scales in the linear r\'egime and  a non-perturbative one for small non-linear scales. He proposes, as an example, to embed LTB solutions into a FLRW universe. 

\section{Super-Hubble fluctuations}

\subsection{Claims}

Some works have been devoted to the study of the effect of energy density fluctuations of wave-length larger than the Hubble radius, generated during inflation, upon cosmic parameters related to the acceleration issue.

Kolb et al (2005a) and Barausse, Matarrese \& Riotto (2005) calculate to second order the corrections to the expansion rate evolution of a matter dominated FLRW universe, in the adiabatic case. They claim that the local Hubble constant and the ``deceleration'' parameter can exhibit a large cosmic variance, depending on the physical conditions prevaling during inflation, with the strongest contribution issued from superhorizon metric perturbations. In this case, the ``deceleration'' parameter has a non-zero probability to be negative. 

The isocurvature case has been analysed by Martineau \& Brandenberger (2005). They compute the backreaction in terms of the Effective Energy Momentum Tensor (EMT) of cosmological perturbations. They claim that, during the matter dominated era, the energy density of the EMT can overcome the cosmological fluid one and that the equation of state of the EMT quickly converges towards that of a cosmological constant.

\subsection{Critics of the above methods}

Ishibashi \& Wald (2006) have criticized the methods employed. 

Considering the analyses by Kolb et al (2005a) and Barausse, Matarrese \& Riotto (2005), they stress that some spatially averaged quantities behaving the same way as in FLRW models with acceleration do not justify a model. A quantity representing the ``scale factor'' may ``accelerate'' without any observable consequence. 

As regards Martineau \& Brandenberger's (2005) work, they emphasize that calculate the second-order EMT and show it has a form and a magnitude similar to that of a cosmological constant is not adequate. A large EMT implies to add contributions to higher perturbative order and therefore the second order method fails.

\subsection{Failure of the superhorizon fluctuation proposals}

Kolb et al (2005b) show that, if adiabaticity holds, super-Hubble perturbations have no impact on local physical observables such as the local expansion rate, therefore invalidating the claims of their own (2005a) article. 

Geshnizjani, Chung \& Afshordi (2005) demonstrate that super-Hubble perturbations yield a correction due to a renormalisation of the local spatial curvature which, by nature, cannot produce an accelerated expansion. 

Flanagan (2005) studies a ``gedanken'' Universe with negligible sub-horizon perturbations and super-Hubble ones equal to those retained by Barausse et al (2005) and Kolb et al (2005a). He shows that, since an acclerated expansion is not driven by superhorizon perturbations in this context, it cannot work in our Universe. 

Hirata \& Seljak (2005) stress that the definition of a deceleration parameter in an inhomogeneous model is tricky. They also give a demonstration of a no-go theorem: cosmological models with irrotational initial conditions (which are generic in standard models of inflation), perturbations at and above the Hubble scale and matter fields matching the strong energy condition cannot explain an accelerating expansion. 

Kolb, Matarrese \& Riotto (2005c) study the isocurvature case, i. e., the proposal of Martineau \& Brandenberger (2005). By solving directly Klein-Gordon equation for the fluctuation of the scalar field representing the density perturbations, they show that super-Hubble fluctuations behave like a relativistic gas. The associated energy density scales alike and may never dominate over the matter. 

These authors have therefore firmly established that superhorizon inhomogeneities cannot mimic the effect of a dark energy component in homogeneous cosmology.

\section{Subhorizon fluctuations}

\subsection{The need}

Buchert (2005) states that to replace the dark energy explanation by kinematical backreaction effects a necessary condition is to show that the cosmos is dominated by strong expansion fluctuations, in constrast to a perturbed Friedmannian state. 

R\"as\"anen (2006) shows that an expansion rate increased by the backreaction of superhorizon inhomogeneities implies a long lasting negative spatial curvature which produces an enhanced Integretated Sachs-Wolf effect. This would increase the amplitude of the low CMB multipoles. Since it would contradict the observations, a more rapid transition from the standard FLRW behavior to the accelerated expansion would be needed. Therefore late-time acceleration must involve spatial curvature and subhorizon perturbations. 

Inhomogeneities likely to solve the cosmological constant and coincidence problems must therefore be of the subhorizon and strong type, which cannot be studied in the framework of perturbation theory.

\subsection{Proposals making use of questioned methods}

Notari (2005) proceeds to second order {\it perturbative corrections} to the FLRW expansion model and obtain {\it an effective Friedmann equation} with {\it an effective dark energy term.}

Moffat (2005) considers a LTB model with an inhomogeneous enhancement at the Hubble horizon scale. Averaging over all {\it the deceleration parameters} measured by observers located in different places, he obtains a cosmic variance for this parameter which can therefore become negative. 

The universe of Nambu \& Tanimoto (2005) seems rather unphysical. It 
contains both a contracting LTB region with positive spatial curvature and another LTB region with negative spatial curvature. A spatial average of this model yields an accelerated expansion as a consequence of the behavior of {\it a deceleration parameter.}
 
\subsection{Proposals aiming at reproduce the observations with exact solutions}

The toy model studied by Mansouri (2005) consists of a LTB patch embedded in a FLRW background with no thin shell at the boundary. The observer can be located anywhere. The luminosity distance-redshift relation of such a model mimics that of a FLRW universe with dark energy.

Alnes, Amarzguioui \& Gron (2006) consider an underdense LTB bubble centered near the observer and surrounded by an Einstein-de Sitter background. They find some subclass of this model which reproduces very accurately (better than the $\Lambda$CDM model) the SNIa data with no need for dark energy and the location of the first peak of the CMB power spectrum, and, approximately, the mass density parameter $\Omega_m$ infered from the mass-to-light ratio measurements made by the 2dF team.

Using different LTB models, Bolejko (2005) finds that realistic matter fluctuations can mimic acceleration at small scales ($z<0.3$). But at larger scales, models fitting the cosmological measurements exhibit peculiar features, possibly due to the spherical symmetry restrictions. The author therefore proposes to try non-symmetrical inhomogeneous cases and, in case of failure, to stick to the cosmological constant explanation.

Vanderveld, Flanagan \& Wasserman (2006) show that many of the proposed models contain a weak singularity at the center, which is the usual location of the observer, and are therefore unphysical. However, some singularity-free models exhibit regions which mimic an accelerating Universe, usually, at low redshifts. 

The last three above articles stress that, to validate inhomogeneous models, these should be able to match all of the cosmological data. In fact, none of the authors who have used LTB models to study the cosmological and coincidence problems have claimed that these models can accurately reproduce the observed universe. These have been considered as mere toy models designed to give a better insight into these issues at low redshifts than the standard $\Lambda$CDM. A spherically symmetric inhomogeneous model can be considered as a non-symmetric one averaged over angular scales, i.e., with less symmetry than the everywhere homogeneous FLRW model. Unfortunately, we know very few exact solutions to Einstein's equations, which could be of use in a cosmological framework.  

\section{Conclusions}

The backreaction of superhorizon fluctuations cannot explain an acceleration of the Universe expansion rate nor mimic a dark energy component. Inhomogeneities likely to solve the cosmological constant and coincidence problems must be of the subhorizon and strong type, which cannot be studied with perturbation methods. 

Exact solutions modelize both strong and weak inhomogeneities. They can solve the cosmological constant problem at low redshift ($z\leq 1$) and, as a consequence, the coincidence problem. Unfortunately, they are likely to exhibit unphysical properties issued from the spherical symmetry of the solutions most frequently used as toy models. To reproduce all the available cosmological observations, it seems that the solution, if any, might be to use non pathological exact inhomogeneous solutions, reproducing the nearby Universe, coupled to nearly homogeneous ones valid up to the scale of the last scattering surface.

\end{document}